\documentclass[onecolumn,showpacs,11pt,nofootinbib]{svjour3}
\usepackage{amssymb,amsmath}

\begin{document}

\title{Effective action for a quantum scalar field in warped spaces}
\author{J. M. Hoff da Silva \and E. L. Mendon\c ca \and E. Scatena}
\institute{Departamento de F\'isica e Qu\'imica, Universidade Estadual Paulista ``J\'ulio de Mesquita Filho'' - UNESP\\
Av. Ariberto Pereira da Cunha 333, Pedregulho, Guaratinguet\'a, SP - Brazil\\
\email{hoff@feg.unesp.br}\and \email{eliasleite@feg.unesp.br}\and
\email{eslley@feg.unesp.br}}
\maketitle

\begin{abstract}
We investigate the one-loop corrections at zero, as well as finite temperature, of a scalar field taking place in a braneworld motived warped background. After to reach a well defined problem, we calculate the effective action with the corresponding quantum corrections to each case. 
\end{abstract}


\section{Introduction}

The subtlety of Quantum Field Theory (QFT) in curved spacetime is well-known. In fact, there are several seminal works elaborating on the many sharp points necessary for the well establishment of such a theory \cite{1,2}. On general grounds it is possible to split the approaches of quantum formulations in curved spacetimes in two branches: extension of the usual formalism by applying and adapting the flat background formulation to the curved case (see \cite{3,4}, just to enumerate some) and, on the other hand, construction of an entire new framework, as the formulation of algebraic quantum field \cite{AL} serve as a prominent example. It is also relevant to stress new approaches outside the perturbative scope \cite{PAS}. In this paper we adopt the former approach, represented by the background-field method \cite{ORI}, and investigate how the quantization upon warped spaces can bring new features for both theories, QFT in curved spaces and non factorizable geometries. 

The application of the usual background-field method for quantization in curved spaces rests upon the (plausible) hypothesis that in a neighbourhood of a given point over the basis manifold, the momentum space can be accessed, at least in some approximation. In this vein, by means of a local momentum space representation, the Minkowski space techniques may be applied. This program can be systematically implemented by means of the so-called coincidence limit \cite{BP}. The novelty in the case to be studied here is that even in this limit there are corrections coming from the warped geometry. 

In fact, our aim here is to investigate the effective action for a quantum scalar field in a five-dimensional warped braneworld background. Obviously, in a given extra dimensional fixed point, the four-dimensional quantization procedure itself is quite usual and shall not be repeated here. Therefore, we shall investigate what (if any) new characteristics can appear from the quantization taking into account the codimension, exploring the warp factor consequences. Having said that, the context is indeed clear. If necessary, however, one can bear in mind the following picture: after the very presentation of the paradigmatic warped braneworld model \cite{RS,LOC}, it was found that is necessary to relax the constraint of standard model fields fixed on the brane \cite{ARK}, giving rise, then, to the universal extra dimensional models, in which all the fields are free to probe the extra dimension. As the universe must be realized, in a manner of speaking, on the brane, the standard model fields must be localized around the brane core. Hence, the quantum field we are interested here can be treated as a quantum fluctuation around the brane. 

As remarked, in the quantization process we make use of the coincidence limit, necessary to engender the homogeneity required to achieve the momentum space representation. The full appreciation of this problem in a background containing a given brane has led to an interesting constraint on the warp factor itself in order to compute the one-loop correction. This criteria is in fully agreement with the more or less recent advances in the characterization of globally hyperbolic spacetimes \cite{DOIS}.    

This paper is organized as follows: in the next section we give an outlook of the basic formalism, indicating the overall procedure to extract the effective lagrangian. In Section 3, we construct the appropriate operator to be inverted, taking into account the specificities coming from the braneworld background. In Section 4 we complete the quantization procedure, generalizing the approach to the finite temperature case in section 5. Finally, in section 6 we conclude. 

\section{Outlook of the formalism}

We start from the scalar field lagrangian defined by\footnote{The possible self-coupling term is not taken into account in the lagrangian, ensuring only perturbatively renormalizable terms in higher dimensions.} 
\begin{equation}
\mathcal{L}\left[\Phi, g_{\mu\nu}\right]=-\frac{1}{2}\Phi\left[\Box +(1-\xi)\xi_{d}R+m^2\right]\Phi,\label{l1}
\end{equation} being $\Box=g^{\mu\nu}\nabla_{\mu}\nabla_{\nu}$ and $\xi_{d}=\frac{1}{4}\frac{(d-2)}{(d-1)}$, where $\xi=1$ denotes the minimal coupling and $\xi=0$ stands for the conformal coupling case. In order to implement the background-field method \cite{dW} we split the field as $\Phi=\hat{\phi}+\phi$, where $\hat{\phi}$ represents the classical background and $\phi$ stands for the quantum fluctuation. The classic configuration dynamics can readily  be read from the usual requirement $\frac{\delta \mathcal{L}}{\delta\Phi}|_{\Phi=\hat{\phi}}=0$, leading to 
\begin{equation}
\bigg(\Box+m^2+(1-\xi)\xi_dR\bigg)\hat{\phi}\equiv \big(\Box+\alpha^2\big)\hat{\phi}=0.\label{eq1}
\end{equation} 
The quantum fluctuation dynamics is ruled by requiring $\frac{\delta \mathcal{L}}{\delta\phi}=0$, which, by means of (\ref{eq1}), gives the same dynamics as the background, Eq. (\ref{eq1}), this time with $\phi$ replacing $\hat{\phi}$. We stress that in the case of considering self-coupling terms in the lagrangian, the dynamics would be slightly different.  

The induced one-loop effects lagrangian, say $\mathcal{L}^{(1)}$, is given by 
\begin{equation}
\exp\Bigg(\frac{i}{\hbar}\int dx \mathcal{L}^{(1)}\Bigg)=N\int (d\phi)\exp\Bigg(-\frac{1}{2}\phi(\Box+\alpha^2)\phi\Bigg),\label{eq2}
\end{equation} where $N$ is a irrelevant normalization factor. Hence, differentiating both sides of (\ref{eq2}) with respect to $\alpha^2$ leads to 
\begin{equation}
\frac{\partial\mathcal{L}^{(1)}}{\partial \alpha^2}=-\frac{1}{2}\frac{\int (d\phi)\phi(x)\phi(x')\exp\{\frac{i}{\hbar}\int dx L\}}{\int (d\phi)\exp\{\frac{i}{\hbar}\int dx L\}}=\lim_{x\rightarrow x'}-\frac{1}{2}\langle \phi(x)\phi(x') \rangle.\label{eq3}
\end{equation} Therefore we have
\begin{equation}
\frac{\partial\mathcal{L}^{(1)}}{\partial \alpha^2}=\lim_{x\rightarrow x'}-\frac{\hbar}{2i}G(x,x'),\label{eq4}
\end{equation} being $G(x,x')$ such that
\begin{equation}
(\Box+\alpha^2)G(x,x')=(-g)^{-1/2}\delta(x,x'),\label{eq5}
\end{equation} and, thus, the one-loop effective lagrangian can be obtained by means of the propagator (computed in the coincidence limit) after integration in $\alpha^2$. The conventions used throughout this paper are such that $x'$ denotes the origin of the two point Green function. A straightforward manipulation of Eq. (\ref{eq5}) leads to
\begin{equation}
H\bar{G}=(-g)^{-1/2}(x)(-g)^{1/2}(x')\delta(x,x'),\label{eq6}
\end{equation} where $H=(-g)^{1/4}(x')[\Box+\alpha^2](-g)^{-1/4}(x)$ and $\bar{G}=(-g)^{1/4}(x)G(x,x')(-g)^{1/4}(x')$. 

As we have remarked, the metric expansion by means of a Riemann coordinate system is, at least, doubtful since the brane breaks the full spacetime diffeomorphism and possibly the necessary homogeneity around a given point near the brane (or at the brane core, in our case). The explicit calculations involving the $H\bar{G}$ operator are lengthy but not particularly difficult. The general idea is just elaborating on the kernel

\begin{eqnarray} 
H\bar{G}=(-g)^{1/4}(x')\Box\bigg((-g)^{-1/4}(x)\bar{G}\bigg)+\alpha^2(-g)^{1/4}(x')(-g)^{-1/4}(x)\bar{G},\label{eq7} 
\end{eqnarray} where the derivative is taken with respect to $x$, {\it i. e.}, 
\begin{eqnarray}
\Box\bigg((-g)^{-1/4}(x)\bar{G}\bigg)=\frac{1}{\sqrt{-g}}\partial_M\Big[(-g)^{1/2}g^{MN}\partial_N[(-g)^{-1/4}\bar{G}]\Big],\label{eq8}
\end{eqnarray} as usual. Thus, after a bit of algebra, Eq. (\ref{eq6}) reads 
\begin{eqnarray} 
g^{AB}\partial_A\partial_B\bar{G}&+&\left.\partial_Ag^{AB}\partial_B\bar{G}+(-g)^{-1/4}(x)\partial_A\Big[(-g)^{-1/2}(x)g^{AB}\partial_B(-g)^{-1/4}(x)\Big]\bar{G}\right.\nonumber\\&+&\left. \alpha^2\bar{G}=(-g)^{1/2}(x')(-g)^{-1/2}(x)\delta(x,x').\right.\label{eq9} 
\end{eqnarray}

Heretofore, we have made no assumptions on the spacetime structure, except for its warped nature and the existence of a brane in the origin $x'$ (which can be assumed as the zero point without loss of generality). Now we shall introduce a few further restrictions to render this study feasible.

\section{Warped braneworld peculiarities}

In order to address to our problem properly, some particularization towards the warped braneworld background are in order. The line element is, thus, given by 
\begin{equation}
ds^2=W(y)^2\eta_{\mu\nu}dx^{\mu}dx^{\nu}+dy^2=g_{AB}dx^{A}dx^{B},\label{metric}
\end{equation} being $\eta_{\mu\nu}=\mathrm{diag}(-1, +1, +1, +1)$, and $g_{AB}=\mathrm{diag}(W(y)\eta_{\mu\nu},1)$, with greek indices running from $\mu=0,1,2,3$. The necessary boundary conditions over $W(y)$, ensuring a well defined background, are 
\begin{eqnarray}
\lim_{y\rightarrow 0}W(y)=1, \quad \lim_{y\rightarrow\pm\infty} W(y)=0,\quad\mathrm{ and} \quad\lim_{y\rightarrow 0}W'(y)=0, \label{CC}
\end{eqnarray} 
where $W'(y)$ means derivative with respect to the extra dimension, and the last condition ensures its maximum in $y=0$. In view of the above line element Eq. (\ref{eq9}) reads 
\begin{eqnarray}
\nonumber&&(\partial_y^2+W^{-2}\Box_4)\bar{G}-\frac{2}{W^{10}(y)}\bigg[2W(y)W''(y)-7(W'(y))^2\bigg]\bar{G}\\&&+\alpha^2\bar{G}=W^{-4}(y)\delta(y)\delta(x),\label{eq10}
\end{eqnarray} 
where the $\bar{G}$ first derivative terms disappear due to fixed coefficient of the $y$-coordinate in the background metric. It is fairly simple to see that the scalar of curvature associated to (\ref{metric}) is 
\begin{equation}
R=-\left[8\frac{W''(y)}{W(y)}+12\frac{(W'(y))^2}{W^2(y)}\right],\label{eq11}
\end{equation} therefore the effective mass term may be reinserted back into Eq. (\ref{eq10}) yielding 
\begin{eqnarray}
\nonumber &&\left(\partial_y^2+W^{-2}\Box_4\right)\bar{G}+m^2\bar{G}+\frac{1}{W^{10}}\left[W''W \left( \frac{3W^8}{2}(\xi-1)-4 \right)   \right.\\&+&\left.\frac{3}{2}W'^2 \left(\frac{3W^8}{2}(\xi-1)+\frac{28}{3}\right)\right]\bar{G}=W^{-4}\delta(y)\delta(x).\label{eq12}
\end{eqnarray}

Now we are in position to face the important question: is the brane thin or thick? As it is well known from the braneworld modeling, the general idea that there is a typical scale below which the standard physics should be modified is incorporated in the thick brane paradigm. We have not, however, completely specified the brane shape so far. In fact the conditions (\ref{CC}) may be applied to both (thick or thin) cases. Another important question concerning the background is the study of hyperbolic operators itself, as the one of Eq. (\ref{eq12}). In order to avoid ill defined scenarios in the construction of the propagator in curved backgrounds, the spacetime must be globally hyperbolic. As one may guess, these two points are also related in this problem. 

In order to solve these problems, we start assuming the brane as an infinitely thin object. We shall see that in this case a contradiction shows up, forcing one to conclude that the brane must be thick. 

By assuming, then, a thin brane we are able to split the spacetime metric, or equivalently the warp factor in our case, and the Green's function itself, along the extra dimension as
\begin{eqnarray}
\bar{G}=\Theta(y)\bar{G}^{+}+\Theta(-y)\bar{G}^{-},\nonumber\\ W=\Theta(y)W^++\Theta(-y)W^-,\label{split}
\end{eqnarray} where $\Theta$ is the usual Heaviside distribution defined by 
\begin{eqnarray}
\Theta(y)=\left \{
\begin{array}{cc}
1, & y>0 \\
0, & y<0 \\
\end{array}
\right.,
\end{eqnarray} obeying the algebra $\Theta^2(y)=\Theta(y)$, $\Theta(y)\Theta(-y)=0$, and $\frac{d\Theta(y)}{dy}=\delta(y)$. This split along the extra dimension is quite useful in order to explore the brane as a region between two different bulk adjacencies \cite{MAC}. The situation expressed by Eqs. (\ref{split}) is clear: we are decomposing the relevant quantities in both sides separated by the brane and the projection on the brane will be performed in a moment. Within the aforementioned algebra, the unity is nothing but the simple sum $\Theta(y)+\Theta(-y)=1$, and therefore it is fairly trivial to see that 
\begin{eqnarray}
W^n=\Theta(y)(W^+)^n+\Theta(-y)(W^-)^n, \hspace{.5cm}\forall n\in\mathbb{Z}.
\end{eqnarray}

In view of our tentative assumption, the brane is understood as a infinitely thin hypersurface orthogonally riddled by geodesics. The derivatives with respect to the extra dimension is given by 
\begin{eqnarray}
W'&=&\Theta(y)\partial_yW^++\Theta(-y)\partial_yW^-,\label{primeira}\\
W''&=&\Theta(y)\partial_y^2W^++\Theta(-y)\partial_y^2W^-+\delta(y)[\partial_yW]n_y,\label{segunda}
\end{eqnarray} where $n_y$ is a unit vector orthogonal to the brane and we denote $[A]=A^+-A^-$, being $A^\pm$ the limit $\lim_{\pm\rightarrow 0}A$, {\it i. e.}, the limit of the $A$ quantity approaching the brane from the side $\pm$. An analogous derivative is respected by the Green's function. By implementing this technique into Eq. (\ref{eq12}) one sees that there is a strong constraint coming from the product $W''W$. In fact, bearing in mind that the product of different distributions is not well defined in the distributional calculus, it is necessary to impose $[\partial_yW]=0$. This is the only way to avoid ill defined scenarios. It turns out, however, that this constraint cannot be fulfilled by an infinitely thin brane. Just to illustrate this point, let us consider the paradigmatic Randall-Sundrum warp factor given by $W=e^{-\kappa|y|}$, being $\kappa$ related to the AdS bulk. It is simple to verify that $(\partial_yW)^+=-\kappa e^{-\kappa y^+}$, while $(\partial_yW)^-=\kappa e^{-\kappa y^-}$. Hence the necessary condition $[\partial_yW]=0$ is not reached.  

As the infinitely brane approach renders the above contradiction, one is forced to conclude that the background must be comprised by a thick brane. This result does not means that quantization over thin branes backgrounds (or over the Randall-Sundrum setup) is wrong. In a plenty of cases of physical relevance, for instance, piecewise continuity of the warp factor is indeed enough. Nevertheless, for the problem we are working on, it is relevant to have a fully smooth background since quantum fluctuations starting from the brane core may cross the brane itself.   

From a different, but complementary, point of view it was recently shown that a given warped spacetime is hyperbolic ({\it i. e.}, it is suitable for a well-defined Cauchy problem) if, and only if, the warp factor is of $C^{\infty}$ class \cite{DOIS}. The previous discussion was, then, an illustrative example of the general formalism developed in Refs. \cite{DOIS}.

\section{Effective lagrangian for a quantum scalar field in warped space}

Having established the brane scenario in which we are working on, we can proceed to seek a solution for the field equations satisfied by the quantum scalar perturbation and, therefore, find the corrections introduced by the one-loop lagrangian $\mathcal{L}^{(1)}$ to the final effective action. 

In order to extract the UV-divergent part of the 1-loop effective action, it is sufficient (and much simpler) just to evaluate the bulk propagator in the vicinity of the brane itself.
Therefore, we can find the associated Green's function by taking the coincidence limit of the Eq. (\ref{eq12}).\footnote{Here and in the rest of this section, $W''=\frac{d^2W}{dy^2}\left.\right|_{y=0}$, since we are in the coincidence limit. Note that even in this case the second derivative of the warp factor is present.} Applying the boundary conditions given by (\ref{CC}), we are left only with the second derivative of the warp factor and do not need to know the full detailed behaviour of $W(y)$ in the interior. With these considerations, we find 
\begin{equation}
\left[\Box_5+\alpha^2\right]\bar{G}_E(z)=\delta(z)\label{eq16},
\end{equation}
with the replacement
\begin{equation}
\alpha^2\rightarrow\alpha^2=\left\{m^2+W''\left[\frac{3}{2}(\xi-1)-4\right]\right\},\label{alpha}
\end{equation}
where we have made the substitution $x-x'=z$.
Here, $\Box _5=\partial_{\tau}^2+\partial_{i}^2+\partial_y^2$ is the Euclidean five-dimensional D'Alembertian obtained via a Wick rotation to the imaginary time $t\rightarrow -i\tau$, and the Euclidean Green's function is defined by $\bar{G}_E(z)=\bar{G}_E(-i\tau,\mathbf{x},y;\tau',\mathbf{x}',y')\equiv\bar{G}(t,\mathbf{x},y;t',\mathbf{x}',y')$.

With this set up we can introduce a momentum space such that the Fourier transform of $\bar{G}_E(z)$ is given by
\begin{equation}
\bar{G}_E(z)=\frac{1}{(2\pi)^5}\int d^5p e^{ipz}G(p),\label{fourier}
\end{equation}
and, using Eq. (\ref{eq16}), the equation satisfied by $G(p)$ is found to be
\begin{equation}
(p^2+\alpha^2)G(p)=1.
\end{equation}
A solution to $G(p)$ can be found upon an inversion of the operator $(p^2+\alpha^2)$ which, in the proper time integral representation \cite{ORI}, assumes the form
\begin{equation}
G(p)=(p^2+\alpha^2)^{-1}=\int_0^{\infty}ds e^{-\alpha^2s}e^{-p^2s}\label{gp}.
\end{equation}
Therefore, using (\ref{fourier}) and (\ref{gp}), the euclideanized Green's function in the configuration space is given by
\begin{eqnarray}
\bar{G}_{E}(z=0)&=&\int_0^{\infty}\frac{ds}{(4\pi s)^{5/2}}e^{-\alpha^2 s},\label{ge}
\end{eqnarray}
and the correction to the total effective lagrangian can be found integrating with respect to $\alpha^2$, reading
\begin{equation}
\mathcal{L}^{(1)}_{E}=-\frac{\hbar}{2(4\pi)^{5/2}}\int_0^{\infty}\frac{ds}{s^{7/2}}e^{-\alpha^2s}.                                                                  
\end{equation}

The above expression for the effective lagrangian is clearly divergent. In order to get a meaningful result, let us perform a small proper-time expansion of the integrand. Since $\alpha^2$ can be split into a mass term, $m^2$, and a term that depends on the warp factor, $U$, such that
\begin{equation}
\alpha^2=m^2+\left\{W''\left[\frac{3}{2}(\xi-1)-4\right]\right\}=m^2+U,\label{alpha2}
\end{equation}
we find
\begin{equation}
\mathcal{L}^{(1)}_{E}=-\frac{\hbar}{2(4\pi)^{5/2}}\int_0^{\infty}\frac{ds}{s^{7/2}}e^{-m^2s}(1-Us+\frac{U^2s^2}{2}-\frac{U^3s^3}{3!}+\frac{U^4s^4}{4!}+\ldots).
\end{equation}
Writing this expansion as $e^{-Us}=\sum_{l=0}^{\infty}a_{l}s^{l}$, and making use of the identity 
\begin{equation}
\Gamma(x)\equiv\int_0^{\infty}ds s^{x-1}e^{-s},\label{gamma}
\end{equation} 
the 1-loop correction to the effective lagrangian can be rewritten as
\begin{equation}
\mathcal{L}^{(1)}_{E}=-\frac{\hbar}{2(4\pi)^{5/2}}\sum_{l=0}^{\infty}a_{l}m^{5-2l}\Gamma\left(l-\frac{5}{2}\right).\label{l1loop}
\end{equation}
In order to better understand where those corrections take place, let us consider a general gravitational action in five dimensions, given by
\begin{equation}
\mathcal{S}_g=\int d^5x \mathcal{L}_0=\int d^5x\sqrt{g}\frac{1}{16\pi G_B}(R-2\Lambda_B),\label{l0}
\end{equation}
where $\Lambda_B$ plays the role of the five-dimensional cosmological constant and $G_B$ is the five-dimensional gravitational constant, all bare quantities \cite{BD}. Therefore, the total effective lagrangian $\mathcal{L}_{eff}$ will be given by the sum of equations (\ref{l1}), (\ref{l1loop}), and $\mathcal{L}_0$ in (\ref{l0}).\footnote{We switch back from the euclideanized form making the substitution $\mathcal{L}^{(1)}_{E}\rightarrow -\mathcal{L}^{(1)}$.}

Expanding the first three terms in the expression (\ref{l1loop}), we find $\mathcal{L}_{eff}=\mathcal{L}+\mathcal{L}_{0}+\mathcal{L}^{(1)}$ to be
\begin{equation}
\mathcal{L}_{eff}=\mathcal{L}+\mathcal{L}_{0}+\frac{m\hbar}{32\pi^2}\left[\frac{4}{15}m^4+\frac{2}{3}m^2U+U^2\right]+\frac{\hbar}{64\pi^{5/2}}\sum_{l=3}^{\infty}a_{l}m^{5-2l}\Gamma\left(l-\frac{5}{2}\right).
\end{equation}

A closer look at equations (\ref{alpha2}) and (\ref{eq11}) shows that $U\propto R$, so we can identify $U=\left[\frac{3}{16}(1-\xi)+\frac{1}{2}\right]R$. In this way, we can rewrite the effective lagrangian in the following form
\begin{equation}
\mathcal{L}_{eff}=\mathcal{L}+\left[\frac{m^5\hbar}{120\pi^2}-\frac{2\Lambda_B}{16\pi G_B}\right]+\left[\frac{m^3\hbar}{48\pi^2}\left(\frac{3}{16}(\xi-1)+\frac{1}{2}\right)+\frac{1}{16\pi G_B}\right]R+\mathcal{O}(R^2).
\end{equation}

As we can see, the first term introduces a correction into the bare cosmological constant, while the second one shifts the gravitational constant. The following terms that appear will introduce small corrections of order $R^2$ and will induce fourth-derivatives of the metric.

It is also interesting to see that even if we set $\Lambda_{B}=0$ in $\mathcal{L}_0$, the corrections arising from the quantum scalar field fluctuations play the role of an effective cosmological constant in five-dimensions with a negative sign, provided a positive definite mass for the scalar field. The relevance of this remark may be understood as follows: in the context of infinitely thin branes there is a sharp relationship between the four-dimensional cosmological constant, $\Lambda_{4D}$, and the five-dimensional counterpart $\Lambda_B$ \cite{JAPAS}. In the case of thick branes the precise analog relation is unknown. It is expected, however, the existence of such a relationship between $\Lambda_{4D}$ and $\Lambda_B$ (otherwise the infinitely thin brane limit would not be possible). In this vein, again in the extreme $\Lambda_B=0$ case, the one-loop corrections would be responsible to set an effective four-dimensional cosmological constant. As an aside remark we stress for the minuteness of the generated $\Lambda_{4D}$.

\section{Finite-temperature corrections}

Taking advantage of the equation (\ref{ge}), we can construct a finite-temperature theory by imposing a periodic condition on the imaginary time $\tau$ according to $\tau\rightarrow\tau+n\beta$, where $\beta=1/k_{B}T$ is the inverse temperature and $k_B$ the Boltzmann constant. Following the procedure outlined in \cite{ORI} and writing the Green's function as a sum in $n$, we obtain
\begin{equation}
G_{\beta}(z,z')=\sum_{n=-\infty}^{\infty}G(x+n\beta u,x'),\quad u=(1,0,0,0,0).
\end{equation}
We can relate the Poisson summation formula and the delta distribution via 
$$\sum_{n=-\infty}^{\infty}e^{ip_0n\beta}=\frac{2\pi}{\beta}\sum_{n=-\infty}^{\infty}\delta\left(p_0-\frac{2\pi n}{\beta}\right),$$
and, for a d-dimensional case, it can be seen that
\begin{equation}
G_{\beta}(z,z')=\frac{1}{(2\pi)^{d-1}\beta}\int d^{d-1}p\int_{0}^{\infty}ds\sum_{n=-\infty}^{\infty}e^{-\alpha^2s} e^{-(2\pi n/\beta)^2s}e^{-|p|^2s}. 
\end{equation}
According to Eq. (\ref{eq4}), the one-loop correction to the effective lagrangian will be given by
\begin{equation}
\frac{\partial\mathcal{L}_{\beta}^{(1)}}{\partial\alpha^2}=\lim_{z\rightarrow z'}\frac{\hbar}{2}G_{\beta}(z,z'),
\end{equation}
and we find, 
\begin{equation}
\mathcal{L}_{\beta}^{(1)}=-\frac{\hbar}{\beta}\frac{\Gamma\left[\frac{1-d}{2}\right]}{2(2\pi)^{d-1}}\sum_{n=-\infty}^{\infty}\left[\alpha^2+\left(\frac{2\pi n}{\beta}\right)^2\right]^{(d-1)/2}.\label{lbeta}
\end{equation}

The correction presented in (\ref{lbeta}) is clearly divergent. In five dimensions $\Gamma\left[\frac{1-d}{2}\right]$ has a sharp divergence, as well as the sum in the free term $\alpha^2$. At this point we shall proceed by manipulating the sum above via usual methods of the regularized zeta function and dimensional regularization. Being willing to accept this procedure is indeed worthwhile, since it is possible to arrive at a finite quantum correction. We start rewriting the sum in (\ref{lbeta}) as

\begin{eqnarray}
\nonumber\sum_{n=-\infty}^{\infty}\left[\alpha^2+\left(\frac{2\pi n}{\beta}\right)^2\right]^{(d-1)/2}&=&\alpha^{d-1}+2\alpha^{d-1}\sum_{n=1}^{\infty}\left(\nu n\right)^{d-1}\left[\left(\frac{1}{\nu n}\right)^2+1\right]^{(d-1)/2},
\end{eqnarray}
with $\nu\equiv\frac{2\pi}{\alpha\beta}$. For a high-temperature expansion ($\alpha\beta\ll 1$), the sum can be expanded as
\begin{eqnarray}
\nonumber\sum_{n=1}^{\infty}(\nu n)^{d-1}\left[\left(\frac{1}{\nu n}\right)^2+1\right]^{(d-1)/2}&\approx& \nu^{\epsilon}\left[\nu^4\zeta(1-d)+\frac{1}{2}(d-1)\nu^2\zeta(3-d)+\frac{1}{8}(d-1)(d-3)\zeta(\epsilon)\right. \\&&\left.+\frac{1}{16}(d-1)(d-3)(d-5)\nu^{-2}\zeta(7-d)+\mathcal{O}(\alpha^3\beta^3)\right],\label{sum2}
\end{eqnarray}
where we have made $\epsilon=d-5$ and $\sum_{k=1}^{\infty}\frac{1}{k^n}=\zeta(n)$ is the regularized zeta function. Combining this result with the expansion for the gamma function given by $\Gamma[\frac{1-d}{2}]=-\frac{1}{d-5}+\left(\frac{3}{4}-\frac{\gamma}{2}\right)+\mathcal{O}(d-5)$ for $d\rightarrow5$ (with $\gamma=0.577...$ being the Euler-Mascheroni constant), and taking the appropriate limit, we find
\begin{eqnarray}
\mathcal{L}_{\beta}^{(1)}=\frac{\hbar}{16\pi^4}\left[-\frac{12}{\beta^5}\zeta(5)+\frac{2\alpha^2}{\beta^3}\zeta(3)-\frac{\alpha^4}{2\beta}\ln{2\pi}+\frac{\alpha^6\beta}{8\pi^2}\zeta(5)+\mathcal{O}(\alpha^7\beta^2)\right].\label{tf}
\end{eqnarray}
The divergent terms proportional to $\frac{1}{d-5}$ combine with those from $\zeta(m-d)$, $m$ being a positive odd integer, yielding a finite result in the limit, whilst the other terms vanish away. Since this is a high-temperature expansion, we can discard higher-order terms and the major contribution will come from the first three. It is worth noting that in this high-temperature expansion the first correction is given solely by the temperature. 

\section{Final remarks}

As can be seen from our results, the appreciation of a quantum scalar field in a warped space braneworld leads to a thick brane background solution in order to define a well-established momentum-space. In fact, as in this problem quantum fluctuations crossing the brane are relevant, it is necessary to have a smooth background. In addition, the warped nature of the metric introduces an effective potential slightly different from what we would expect from an ordinary five-dimensional Einstein-Hilbert gravity with a scalar field, for the zero-temperature theory.

For the finite temperature case, it is also interesting to observe that if we take the limit of $\alpha\beta\rightarrow0$ in Eq. (\ref{tf}) for a very high-temperature expansion, the first terms will diverge and the one-loop correction is no longer sufficient, being necessary higher-loop contributions to tame the infinities that appear.

To sum up, the method adopted in this work has shown that the warp factor in the quantization procedure has a twofold role: on the one hand it is responsible for specific consequences in the one-loop quantum corrections. On the other hand, its adequate functional form is essential to yield a well defined problem.  

\section*{Acknowledgements}

The authors were benefited from useful conversation with Profs. M. B. Hott and C. A. Bonin. JMHS thanks to CNPq 308623/2012-6; 445385/2014-6 for financial support, ELM thanks to CNPq 449806/2014-6, and ES thanks to CAPES/PNPD grant.

\end{document}